\documentclass[preprint,review,3p,12pt]{elsarticle}

\usepackage{amsmath}
\usepackage{graphics}
\usepackage{color}
\usepackage{array}
\usepackage{booktabs}
\usepackage{rotating}

\newcolumntype{P}[2]{%
  >{\begin{turn}{#1}\begin{minipage}{#2}\small\raggedright\hspace{0pt}}l%
  <{\end{minipage}\end{turn}}%
}

\journal{Physica A}

\begin{document}

\begin{frontmatter}

\title{Empirical analysis on the connection between power-law distributions and allometries for urban indicators}
\author[a,c]{Luiz G. A. Alves}
\ead{lgaalves@dfi.uem.br} 
\author[a,b,c]{Haroldo V. Ribeiro}
\ead{hvr@dfi.uem.br}
\author[a,c]{Ervin K. Lenzi} 
\author[a,c]{Renio S. Mendes}  

\address[a]{Departamento de F\'isica, Universidade Estadual de Maring\'a - Maring\'a, PR 87020-900, Brazil}
\address[b]{Departamento de F\'isica, Universidade Tecnol\'ogica Federal do Paran\'a - Apucarana, PR 86812-460, Brazil}
\address[c]{National Institute of Science and Technology for Complex Systems, CNPq - Rio de Janeiro, RJ 22290-180, Brazil}

\begin{abstract}
We report on the existing connection between power-law distributions and allometries. As it was first reported in [PLoS ONE 7, e40393 (2012)] for the relationship between homicides and population, when these urban indicators present asymptotic power-law distributions, they can also display specific allometries among themselves. Here, we present an extensive characterization of this connection when considering all possible pairs of relationships from twelve urban indicators of Brazilian cities (such as child labor, illiteracy, income, sanitation and unemployment). { Our analysis} reveals that all our urban indicators are asymptotically distributed as power laws and that the proposed connection also holds for our data when the allometric relationship displays enough correlations.  We have also found that not all allometric relationships are independent and that they can be understood as a consequence of the allometric relationship between the urban indicator and the population size. We further show that the residuals fluctuations surrounding the allometries are characterized by an almost constant variance and log-normal distributions.
\end{abstract}

\end{frontmatter}
\section*{Introduction}
There has been a remarkable and growing interest in investigating social systems through the framework of statistical physics over the last decades~\cite{Boccaletti,Castellano,Galam,Vespignani,Conte,Takahashi}. The majority of these studies are however focused on models and proprieties that often resemble those of phase transitions~\cite{Castellano}. Despite its evident importance, considerably less and uneven attention has been paid towards working out empirical data related to social systems. For instance, empirical-driven investigations on financial markets~\cite{Sornette,Mantegna} are much more abundant than those related to urban systems. Furthermore, since more than a half of the human population lives in urban areas~\cite{Crane,WRI}, it is crucial to identify and understand patterns of urban systems.  

Among several properties of urban systems, there are two which are remarkable and ubiquitous: the emergence of asymptotic power-law distributions and allometries for wide range of variables that somehow characterize the urban systems (urban indicators or metrics). On one hand, population~\cite{Marsili,Batty}, building sizes~\cite{Batty}, personal fortunes~\cite{Newman}, firm sizes~\cite{Axtell}, number of patents~\cite{ONeale}, and other indicators have been found to asymptotically follow power-law distributions. On the other hand, allometries with the population size have been reported for crime~\cite{Bettencourt,Alves,Alves2}, suicide~\cite{Melo}, several urban metrics including patents, gasoline stations, gross domestic product~\cite{Bettencourt2,Arbesman,Bettencourt3}, number of election candidates~\cite{Mantovani}, party memberships~\cite{Mantovani2}, among others. These allometries were recently modeled by Bettencourt~\cite{Bettencourt4} via a small set of simple and locally-based principles. However, it was only recently that the connection between these two features was elucidated by Gomez-Lievano \textit{et al.}~\cite{Lievano}. In their work, they have shown, for the relationship between homicides and population, that when both urban indicators are asymptotically distributed as power laws, these urban indicators can also exhibit a particular allometric relationship. More specifically, the exponent of the allometry can be fully determined from the exponents of the power-law distributions. Here, we aim to empirically extend these results through an extensive characterization of all possible pairwise relationships from 12 urban indicators of Brazilian cities. {Our results show that the connection proposed by Gomez-Lievano \textit{et al.}  also holds for our data when there is enough correlation in the allometric relationship. We also discuss that not all allometric exponents are independent and that, in fact, the allometries between pairs of urban indicators can be understood as a consequence of the allometric relationship between the urban indicator and the population size.

This work is organized as follows. We first describe our database. We next investigate the hypothesis that these urban indicators are asymptotically distributed as power laws by employing the statistical procedure proposed by Clauset \textit{et al.}~\cite{Clauset}. Then, we characterize all possible allometric relationships between pairs of urban indicators by accounting for the degree of correlation between them. We thus discuss that when enough correlation in the allometry exists, the allometric exponent can be determined from the power-law exponents of the distributions of urban indicators. We also discuss that the constant behavior of the variance of the fluctuations surrounding the allometries and the log-normal distribution of these residuals are consistent conditions for the applicably of the connection between power laws and allometries. Finally, we present a summary of our findings.

\section*{Data Presentation}
We have accessed data of Brazilian cities in the year of 2000 made freely available by the Brazil's public healthcare system --- DATASUS~\cite{datasus}. In particular, for our analysis, { we select 2862 cities (about $51\%$ of Brazilian cities) for which all values of the urban indicators were available}. The database is thus composed of twelve urban indicators $Y_i$ at city level. They are: total population ($i=0$), 
number of cases of child labor ($i=1$), 
population older than 60 years ($i=2$), 
female population ($i=3$), 
gross domestic product --- GDP ($i=4$), 
GDP per capita ($i=5$), 
number of homicides ($i = 6$), 
number of illiterate older than 15 years ($i=7$), 
average family income ($i=8$), 
male population ($i=9$), 
number of sanitation facilities ($i=10$), 
and number of unemployed older than 16 year ($i=11$).
More details about these urban indicators can be found in the Refs.~\cite{datasus,Alves2}. It is worth noting that, despite there being more than one definition for the concept of city~\cite{Angel}, we here have considered that cities are the smallest administrative units with a local government, and it is beyond the scope of this work to discuss the role of other definitions.

\section*{Methods and Results}
We start our investigation by testing the hypothesis that the 12 urban indicators are distributed according to asymptotic power laws. In order to do so, we first evaluate the cumulative distribution functions for all urban indicators. As shown in fig.~\ref{fig1}, { the shapes of the distributions can be approximated by asymptotic power laws} [$P_i(Y_i) \propto Y_i^{- \alpha_i}$], evidenced here by the linear behavior (in log-log plots) of the distributions for large values of urban indicators. To strengthen this result, we employ the statistical procedure proposed by Clauset \textit{et al.}~\cite{Clauset}, which is a systematic way for statistically testing empirical power-law distributions. Due the discrete nature of most of our urban indicators, we have considered the discrete version of the procedure. In this case, we apply the maximum-likelihood fitting procedure by considering the probability distribution function
\begin{equation}\label{eq:zipf}
P_i(Y_i) = \frac{Y_i^{-\alpha_i}}{\zeta(\alpha_i,Y_{i, \text{min}})},
\end{equation}
where $\zeta(\alpha_i,Y_{i, \text{min}})=\sum_{n=0}^{\infty}(Y_{i, \text{min}}+n)^{-\alpha_i}$ is the Hurwitz zeta function (it ensures the normalization of $P_i(Y_i)$ when $Y_i$ only takes discrete values), $Y_{i,\text{min}}$ is a parameter representing the beginning of the power-law regime, and $\alpha_i$ is the power-law exponent. After finding the best fit parameters, we evaluate the goodness-of-fit via the Cram\'er-von Mises test.

We have thus applied Clauset \textit{et al.}'s procedure to our data and the results, that is, the values of $\alpha_i$ and $Y_{i,\text{min}}$ as well as the $p$-value of the Cram\'er-von Mises test are also shown in fig.~\ref{fig1}. Notice that we cannot reject the power-law hypothesis at a confidence level of 99\% for nearly all urban indicators. The only exception is the number of homicides, for which the $p$-value is $0.002$. We observe, however, that the Cram\'er-von Mises statistic is dominated by the extreme values of the number of homicides, where we note a cut-off-like behavior. In this case, after removing the cities with the 10 largest values, the power-law hypothesis cannot be rejected. { It is worth noting that, for testing the multiple hypothesis that all urban indicators are asymptotically power-law distributed, we need to consider (for instance) the Bonferroni correction~\cite{Miller}. For our case, it implies in assuming a significance level of $0.05/12$ for each hypothesis testing (for confidence level of 95\%). Still, we cannot reject the asymptotic power-law hypothesis for all urban indicators (except by number of homicides). Despite the good agreement, we note that the power-law fits hold only for the tails of the empirical distributions, that is, for a small fraction of the entire dataset. In order to describe not only the tails of these distributions, some authors have employed other functional forms such as the log-normal~\cite{Eeckhout} and the stretched exponential~\cite{Sornette2} distributions. For our data, these two distributions have failed to describe the entire dataset. Because our main goal is to investigate the connection between power laws and allometries, we will concentrate on the tails of these distributions, for which we have verified that the 12 urban indicators can be modeled by asymptotic power laws,} confirming our initial hypothesis and also in agreement with previous reports based on other countries data~\cite{Bettencourt2,Arbesman,Bettencourt3}.

\begin{figure*}[!ht]
\centering
\includegraphics[scale=0.37]{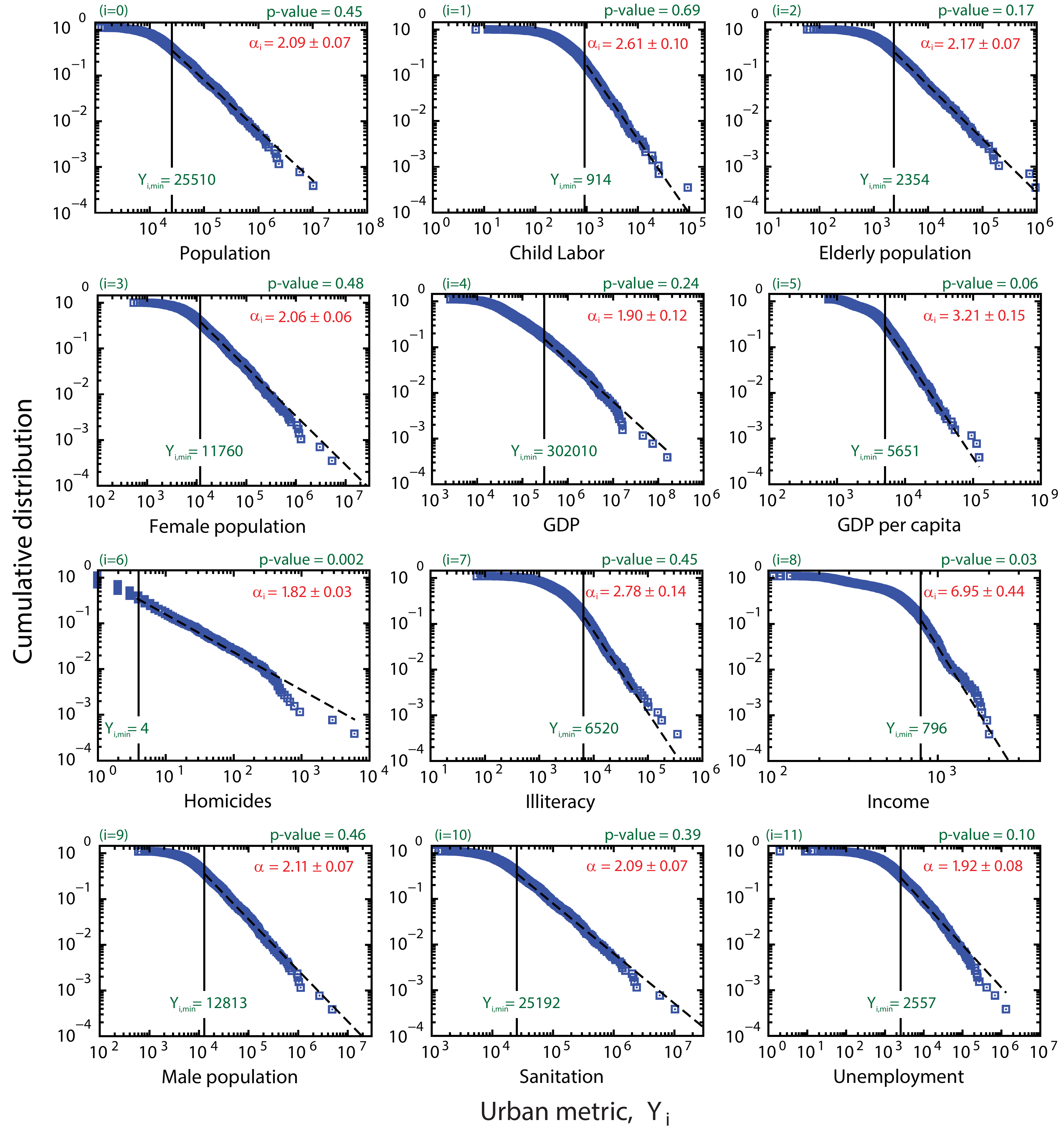}
\caption{Cumulative distributions of 12 urban indicators of Brazilian cities in the year 2000. We note that all distributions display asymptotic power-law decays which are well described by power-law functions $P_i(Y_i) \sim Y_i^{-\alpha_i}$ for $Y_i \geq Y_{i,\text{min}}$ (dashed lines). In each case, the parameters $\alpha_i$ and $Y_{i,\text{min}}$ (shown in the plots) were obtained via maximum-likelihood fits. We also observe that the $p$-values  of the Cram\'er-von Mises tests (shown in the plots) indicate that we cannot reject the power-law hypothesis at a confidence level of 99\% for all indicators, except by the number of homicides (see discussion in the main text).}
\label{fig1}
\end{figure*}

We now focus on the question of whether the power-law distributions for urban indicators imply in allometric relationships between pairs of indicators. As we have mentioned, this question was { first} posed in the work of Gomez-Lievano \textit{et al.}~\cite{Lievano} when investigating the relationships between population size and number of homicides. The following analytical developments are only a revision of the results present in their article. Gomez-Lievano \textit{et al.} proposed that if the shape for the allometry is well established, we can write
\begin{equation}\label{eq:p}
P_i(Y_i)=\sum_{Y_j=Y_{j}^{*}}^{\infty} P_{i,j}(Y_i \mid Y_j)P_j(Y_j)\,,
\end{equation}
where $P_i(Y_i)$ is the probability distribution of the urban indicator $Y_i$ and $P_{i,j}(Y_i\mid Y_j)$ is the conditional probability representing the fluctuations surrounding the allometry between the indicators $Y_i$ and $Y_j$. We already know that $P_i(Y_i)$ has the power-law form of the eq.~(\ref{eq:zipf}) and following empirical findings (details will be provided soon on), we will assume that the conditional probability is a log-normal distribution
\begin{equation}
P_{i,j}(Y_i\mid Y_j)=\frac{1}{Y_i\sqrt{2\pi\,\sigma_{i,j}^2(Y_j)}}\exp\left\{-\frac{[\ln Y_i - \mu_{i,j}({Y_j})]^2}{2\,\sigma_{i,j}^2(Y_j)}\right\}\,,
\end{equation}
in which $\mu_{i,j}({Y_j})$ and $\sigma_{i,j}^2(Y_j)$ represent possible dependences of the log-normal parameters on the indicator $Y_j$. In particular, because we are also assuming that there exists an allometry between $Y_i$ and $Y_j$ (that is, $Y_i\propto Y_j^{\beta_{i,j}}$), the functional form of $\mu_{i,j}({Y_j})$ is 
\begin{equation}
\mu_{i,j}({Y_j}) = \ln[\mathcal A_{i,j}\, Y_j^{\beta_{i,j}}]\,,
\end{equation}
if we consider the standard deviation not dependent on $Y_j$ (that is, $\sigma_{i,j}^2(Y_j)=\sigma_{i,j}^2$). 

After considering all the previous assumptions, we can employ eq.~(\ref{eq:p}) by replacing the sum by an integral and considering $Y_{j}^{*}$ sufficiently small. In fact, after some calculations (see methods section of Ref.~\cite{Lievano} for details), we have
\begin{equation}
P_i(Y_i)\propto Y_i^{-\frac{(\alpha_j-1)}{\beta_{i,j}}-1}\,.
\end{equation}
Now, remembering that $P_i(Y_i)\sim Y_i^{-\alpha_i}$, we can write the relationship between the allometric exponent ($\beta_{i,j}$) and the power-law exponents ($\alpha_i$ and $\alpha_j$) as
\begin{equation}\label{eq:beta}
\beta_{i,j}=\frac{\alpha_j-1}{\alpha_i-1}\,.
\end{equation}
It is worth noting that the same relationship could have been obtained by using $P_i(Y_i)dY_i=P_j(Y_j)dY_j$. However, when doing this calculation we also have to assume that the allometry between $Y_i$ and $Y_j$ is an exact expression (not displaying any randomness), which is not the case as we shall see. 

The developments of Gomez-Lievano \textit{et al.}~\cite{Lievano} thus analytically prove that when the urban indicators present power-law distributions, they can also display allometric relationships with particular exponents. In order to empirically test this result, we investigate all possible allometries between pairs of urban indicators in our data. Specifically, we have analyzed all these relationships by considering the logarithm of the urban indicators and by adjusting (via ordinary least square method) the linear function
\begin{equation}
\log_{10} Y_i = A_{i,j}+\beta_{i,j}^{(f)} \log_{10} Y_j
\end{equation}
to all of them. Here $A_{i,j}=\log_{10}\mathcal A_{i,j}$ is an empirical constant and $\beta_{i,j}^{(f)}$ is the empirical value of the allometric exponent (obtained after fitting the allometry). Figure~\ref{fig2} shows examples of allometries between the urban indicators population size, number of homicides and number of cases of child labor. Towards finding the values of $\beta_{i,j}^{(f)}$, we have applied a cut-off in the abscissas axis considering only the values of $Y_j$ that are larger than $Y_{j,\text{min}}$ (the beginning of the power-law regime in the distribution of $Y_j$, see fig.~\ref{fig1}). The blue dots in fig.~\ref{fig2} are the points obeying this condition and the dashed lines represent the linear functions adjusted to the data. We repeat this procedure to all possible allometries and fig.~\ref{fig3} shows a matrix plot with all the values of $\beta_{i,j}^{(f)}$. We have further characterized the quality of these allometries by calculating the Pearson correlation coefficient $\rho_{i,j}$ { for the linearized allometric relationships (that is, $\log_{10} Y_i$ versus $\log_{10} Y_j$)}, which is displayed in fig.~\ref{fig3} through the color code. We note that the majority of the relationships present $\rho_{i,j}$ values larger than 0.5 ($\approx\!70\%$). However, we also observe two systematic exceptions: the allometries with urban indicators GDP per capita and income. { The reason for this deviant behavior is related to the fact that these two indicators are defined per capita and per family. This result also suggests that the main source of correlation between two indicators $Y_i$ and $Y_j$ ($i\neq j\neq0$) comes from their relationships with the population size $Y_0$ (we will investigate this hypothesis soon).}

\begin{figure}[!ht]
\centering
\includegraphics[scale=0.45]{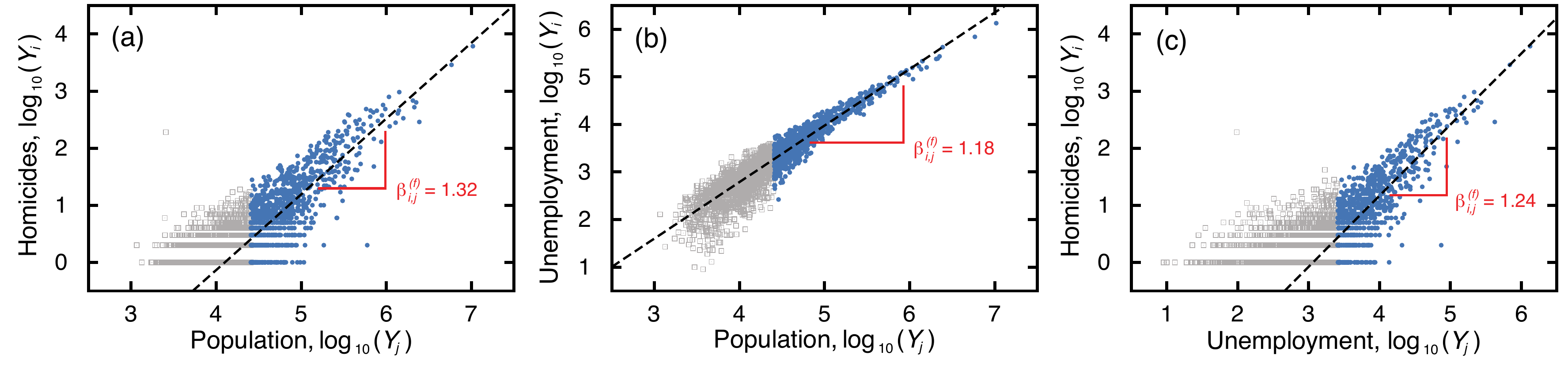}
\caption{Examples of allometries between urban indicators: (a) homicides versus population, (b) child labor versus population and (c) homicides versus child labor. The markers are base-10 logarithm of the values of the urban indicators $Y_i$ versus $Y_j$ for each city. The blue dots represent those cities which have $Y_j \geq Y_{j,\text{min}}$ and gray squares are those for which $Y_j < Y_{j,\text{min}}$ ($Y_{j,\text{min}}$ is the one obtained in fig.~\ref{fig1}). The dashed lines are ordinary linear least square fits to the points obeying the condition $Y_j \geq Y_{j,\text{min}}$ and the slope of each line is equal to the allometric exponent $\beta_{i,j}^{(f)}$ (shown in the plot).}
\label{fig2}
\end{figure}

Now, we can quantitatively compare the values of $\beta_{i,j}^{(f)}$ (estimated by fitting the allometries) to the values of $\beta_{i,j}$ (analytically obtained from eq.~(\ref{eq:beta})). Figure~\ref{fig4}(a) shows a scatter plot of $\beta_{i,j}^{(f)}$ versus $\beta_{i,j}$ for allometries with Pearson correlation $\rho_{i,j}$ larger than 0.5. We observe that a linear function (with no additive constant) describes quite well the relation between $\beta_{i,j}^{(f)}$ and $\beta_{i,j}$, specially if we take the standard errors of the values of $\beta_{i,j}^{(f)}$ (shaded area) into account. We have also observed that the quality of this linear relationship deteriorates when we start considering allometries with smaller values of $\rho_{i,j}$. Thus, in addition to be distributed as power laws, the indicators $Y_i$ and $Y_j$ should also present a good quality allometry (large value of $\rho_{i,j}$) in order for the relationship of eq.~(\ref{eq:beta}) to be valid.

\begin{figure}[!ht]
\centering
\includegraphics[scale=0.45]{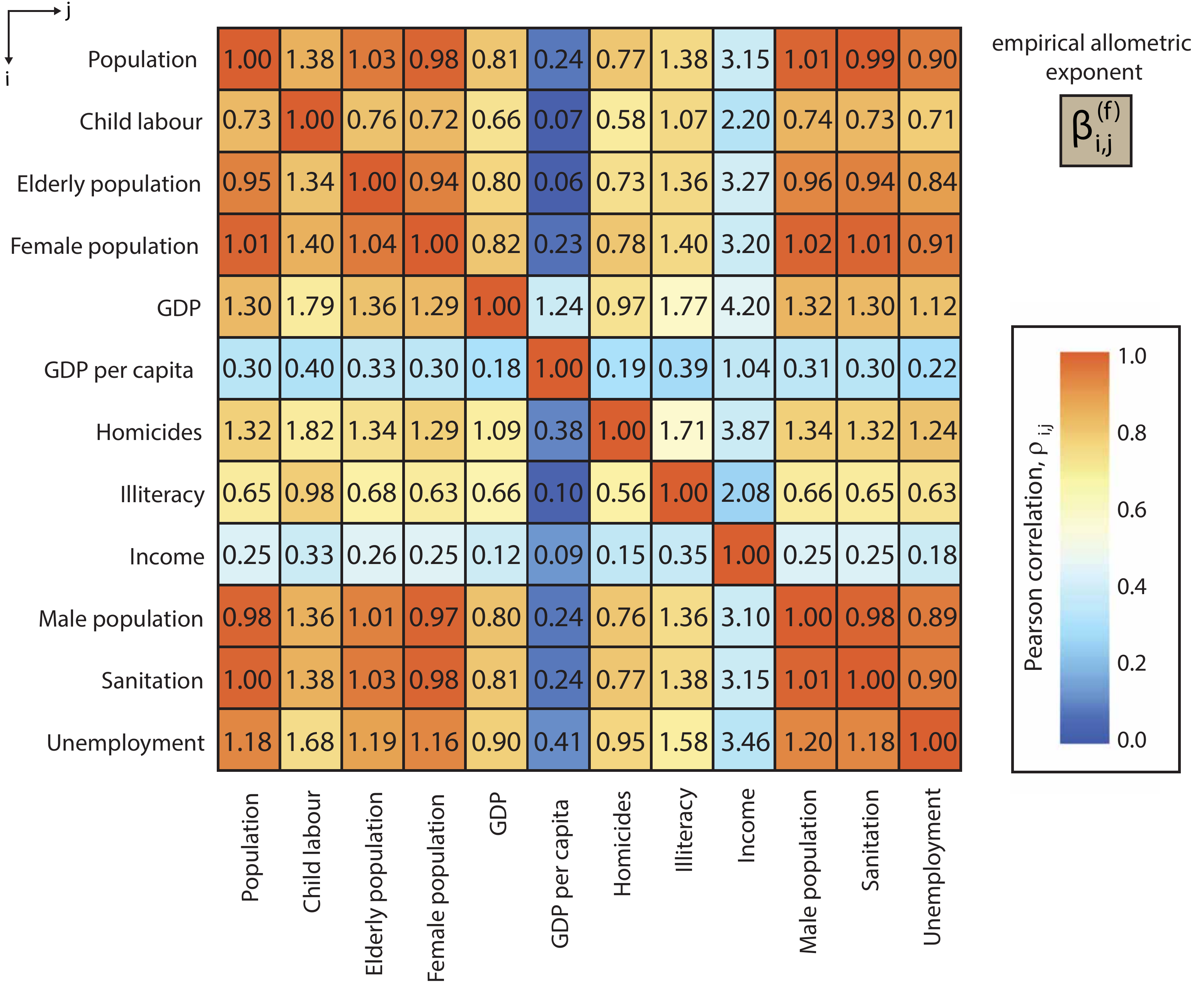}
\caption{Empirical values of the allometric exponents $\beta_{i,j}^{(f)}$ obtained from all possible pairs of relationship { (in logarithmic scale)} between our 12 urban indicators. This matrix plot shows the value of $\beta_{i,j}^{(f)}$, where the row captions represent the case in which the urban indicator is employed as the dependent variable ($y$ axis) and column captions represent the case in which the urban indicator is employed as the independent variable ($x$ axis). The color code shows the value of Pearson correlation coefficient $\rho_{i,j}$ for each allometry.}
\label{fig3}
\end{figure}

It is worth noting that although we have considered all power-law exponents $\beta_{i,j}^{(f)}$ independent, they are actually dependent on each other. Firstly, the allometric exponent of the relationship between $Y_i$ and $Y_j$ is supposed to be the inverse of that obtained for the relationship between $Y_j$ and $Y_i$. For the empirical values, we observe small deviations in this relationship due to the different cut-offs applied to the abscissa axis. However, when taking the confidence intervals for the values of $\beta_{i,j}^{(f)}$ into account, we have verified that this inverse relationship between the values of allometric exponents holds in most of the cases. Secondly and, importantly, when we have that $Y_i\sim Y_k^{\beta_{i,k}^{(f)}}$ and $Y_j\sim Y_k^{\beta_{j,k}^{(f)}}$ directly follow that $Y_i\sim Y_j^{\beta_{i,j}^{(f)}}$ with $\beta_{i,j}^{(f)} = \beta_{i,k}^{(f)}/\beta_{j,k}^{(f)}$ (for noiseless relationships). This result implies that the allometries between all pairs of indicators $Y_i$ ($i\neq0$) can be understood as a consequence of the allometric relationship between the indicator $Y_i$ and the population size $Y_0$. Nevertheless, noisy allometric relationships and the fact that $P_i(Y_i)$ is not a perfect power-law may have a nontrivial role in the empirical values of $\beta_{i,j}$. In fig.~\ref{fig4}(b), we have empirically tested the relationship $\beta_{i,j}^{(f)} = \beta_{i,0}^{(f)}/\beta_{j,0}^{(f)}$ and the results show that it holds for allometries with $\rho_{i,j}$ larger than 0.5. Also in this case, the quality of linear relationship between $\beta_{i,j}^{(f)}$ and $\beta_{i,0}^{(f)}/\beta_{j,0}^{(f)}$ deteriorates when we start considering allometries with smaller values of $\rho_{i,j}$. 

\begin{figure}[!ht]
\centering
\includegraphics[scale=0.6]{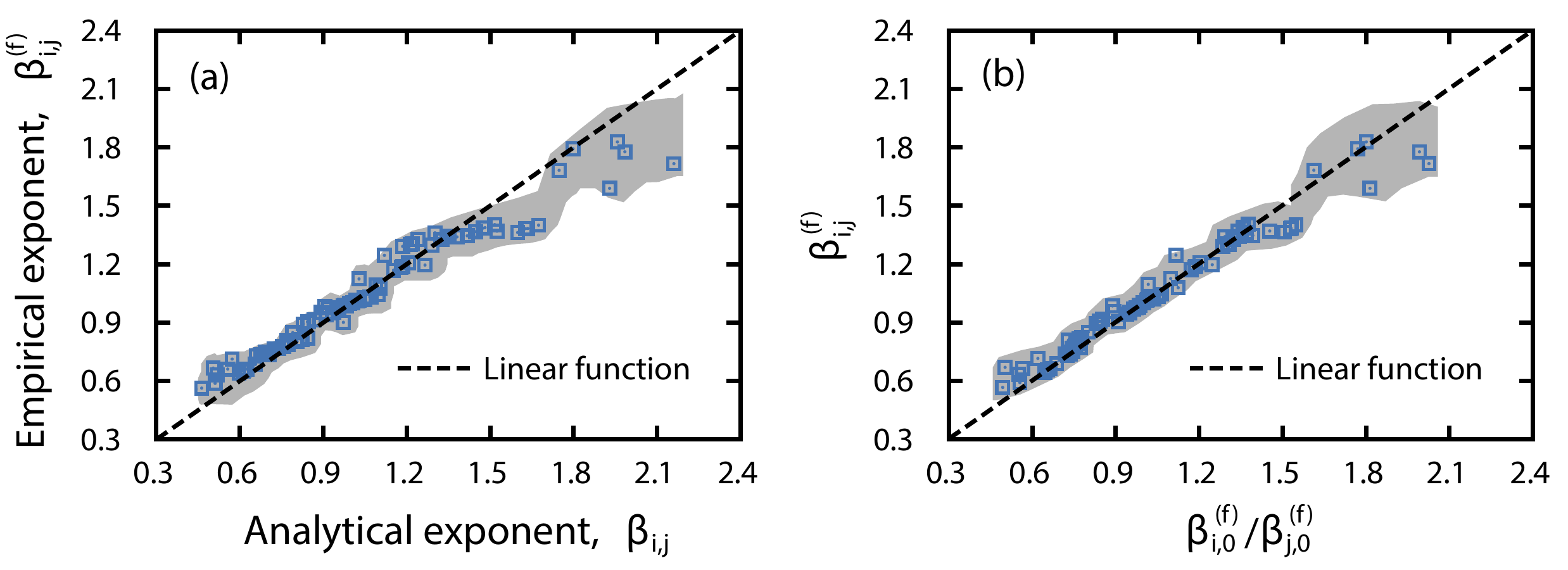}
\caption{(a) Comparison between the empirical values of the allometric exponents $\beta_{i,j}^{(f)}$ (obtained via linear fits) and the analytical ones $\beta_{i,j}$ (obtained from eq.~(\ref{eq:beta})). The squares are the values of $\beta_{i,j}^{(f)}$ versus $\beta_{i,j}$ where the Pearson correlation coefficient $\rho_{i,j}$ characterizing the allometry is larger than $0.5$. (b) Empirical verification of the linear relationship between $\beta_{i,j}^{(f)}$ and $\beta_{i,0}^{(f)}/\beta_{j,0}^{(f)}$ (squares) when considering $\rho_{i,j}\geq 0.5$. In both plots, the shaded areas represent the standard errors of $\beta_{i,j}^{(f)}$ after smoothing the data with a moving average filter of window size 5 and the dashed lines are linear functions with no additive constant.}
\label{fig4}
\end{figure}

{ The fact that the allometry between $Y_i$ and $Y_j$ ($i,j\neq0$) being a consequence of their allometric relationships with the population size $Y_0$ does not necessarily mean that the indicator $Y_j$ has no explanatory potential for describing $Y_i$. In order to test this possibility, we have adjusted the following generalized linear model:
\begin{equation}
\log_{10} Y_i = a_{i,j} + b_{i,0} \log_{10} Y_0 + b_{i,j} (\log_{10} Y_j)/ (\log_{10} Y_0)^{\beta_{i,j}^{(f)}} \,,
\end{equation}
to our data, considering all possible pairs of relationships between $\log_{10} Y_i$ and $\log_{10} Y_j$, but excluding the indicators elderly population, female population, male population and sanitation, since they form an almost perfect linear relationship with the population. Thus, every $b_{i,j}$ statistically different from zero indicates that $\log_{10} Y_j$ presents an explanatory potential for describing $\log_{10} Y_i$, which is not related to its own relationship with $\log_{10} Y_0$. In table~\ref{tab:bs}, we show the regression results for which we cannot reject the hypothesis of the coefficient $b_{i,j}$ being different from zero (at a confidence level of 99\%). Note that, from the 49 possible pairs of relationships, 24 present a $b_{i,j}$ statistically significant. Thus, the noise surrounding the allometries with the population actually contains additional information that can be used for explaining the $Y_i$ in function of $Y_j$. We have made similar considerations when considering the distance to the power laws for unveiling relationships between crime and urban indicators~\cite{Alves2}. }



\linespread{1.}
\begin{table}[!ht]
  \centering
  \caption{Values of the linear coefficients $a_{i,j}$ (intercept), $b_{i,0}$ (population) and $b_{i,j}$ (indicator) obtained via ordinary least-squares fits with a correction to heteroskedasticity. Here we show only the regressions where $b_{i,j}$ is statistically different from zero.  
  }
  \vspace{0.1cm}
  \label{tab:bs}
  \begin{tabular}{@{}lccccccccc@{}}
  \toprule
  & \multicolumn{1}{P{79}{2.9cm}}{Intercept} &
    \multicolumn{1}{P{79}{2.9cm}}{Population} &
    \multicolumn{1}{P{79}{2.9cm}}{Child labor} &
    \multicolumn{1}{P{79}{2.9cm}}{GDP} &
    \multicolumn{1}{P{79}{2.9cm}}{GDP per capita} & 
    \multicolumn{1}{P{79}{2.9cm}}{Homicides} & 
    \multicolumn{1}{P{79}{2.9cm}}{Illiteracy} &
    \multicolumn{1}{P{79}{2.9cm}}{Income} &
    \multicolumn{1}{P{79}{2.9cm}}{Unemployment} 
    \\
  \midrule
  {\bf Child labor} & -1.99 & 0.69 & -- & -- & -- & -- & 1.26 & -- & --  \\
    & 0.95 & 0.84 & -- & -- & -- & -- & -- & -1.05 & -- \\
    & 0.74 & 0.86 & -- & -- & -- & -- & -- & -- & -3.26  \\
  \midrule
  {\bf GDP} & -4.16 & 1.21 & -- & -- & 1.68 & -- & -- & -- & --  \\
    & 0.80 & 0.97 & -- & -- & -- & 1.94 & -- & -- & -- \\
    & 1.77 & 1.27 & -- & -- & -- & -- & -1.69 & -- & -- \\
    & -2.17 & 1.05 & -- & -- & -- & -- & -- & -- & 4.89 \\
  \midrule  
  {\bf GDP per capita} & -5.22 & 0.50 & -- & 9.16 & -- & -- & -- & -- & -- \\
    & 3.80 & -0.03 & -- & -- & -- & 1.94 & -- & -- & -- \\
    & 4.77 & 0.27 & -- & -- & -- & -- & -1.69 & -- & -- \\      
    & 0.83 & 0.05 & -- & -- & -- & -- & -- & -- & 4.89 \\      
  \midrule  
  {\bf Homicides} & -10.07 & 1.45 & -- & 5.57 & -- & -- & -- & -- & -- \\
    & -7.51 & 1.28 & -- & -- & 0.98 & -- & -- & -- & -- \\
    & -9.81 & 1.11 & -- & -- & -- & -- & -- & -- & 9.37 \\
  \midrule  
  {\bf Illiteracy} & -0.92 & 0.72 & 1.05 & -- & -- & -- & -- & -- & -- \\
    & 1.45 & 0.74 & -- & -1.96 & -- & -- & -- & -- & -- \\
    & 0.66 & 0.80 & -- & -- & -0.38 & -- & -- & -- & -- \\
    & 1.78 & 0.84 & -- & -- & -- & -- & -- & -1.17 & -- \\
  \midrule  
  {\bf Income} & 2.55 & 0.24 & -0.86 & -- & -- & -- & -- & -- & -- \\
    & 3.00 & 0.27 & -- & -- & -- & -- & -1.09 & -- & -- \\
  \midrule  
  {\bf Unemployment} & -1.05 & 1.13 & -0.70 & -- & -- & -- & -- & -- & -- \\
    & -2.92 & 1.15 & -- & 1.61 & -- & -- & -- & -- & -- \\
    & -2.21 & 1.09 & -- & -- & 0.30 & -- & -- & -- & -- \\
    & -1.16 & 0.99 & -- & -- & -- & 1.21 & -- & -- & -- \\
  \bottomrule
  \end{tabular}
\end{table}

\linespread{1.5}

We now address the question of the fluctuations surrounding the allometries aiming to verify the two remaining hypotheses in the calculations of Gomez-Lievano \textit{et al.}: the constant behavior of the log-normal parameter $\sigma_i^2(Y_i)$ and the log-normal distribution of fluctuations. In order to access these hypotheses, we have binned the linearized allometric relationships ($\log_{10} Y_i$ versus $\log_{10} Y_j$ with $Y_j>Y_{j,\text{min}}$) in $w$ equally spaced windows, and for each one, we evaluate the variance $\sigma^2_{{i,j}_w}$. Notice that in this log-log scale the value of variance 
$\sigma^2_{{i,j}_w}$ is approximately equal to the value of log-normal parameter $\sigma_{i,j}^2(Y_j)$; therefore, verifying that $\sigma^2_{{i,j}_w}$ is not dependent on $\log_{10} Y_j$ is equivalent to showing that $\sigma_{i,j}^2(Y_j)$ is constant. Figure~\ref{fig5}(a) shows the behavior of $\sigma^2_{{i,j}_w}$ on the window average value of $\log_{10} Y_j$. We observe that there is no clear dependence in these relationships and that the behavior can be approximated (roughly for some allometries) by a constant function, where the plateau value changes from indicator to indicator. Finally, to study the log-normal hypothesis, we evaluate the normalized fluctuations surrounding the linearized allometric relationships, that is,
\begin{equation}
\xi_{i,j} = \frac{\log_{10} Y_i - \langle \log_{10} Y_i \rangle_{w} }{\sigma_{{i,j}_w}}\,,
\end{equation}
where $\langle \log_{10} Y_i \rangle_{w}$ stands for the window average value of $\log_{10} Y_i$. Figure~\ref{fig5}(b) shows the cumulative distribution of $\xi_{i,j}$ for all our urban indicators. We observe that these distributions are in well agreement with the standard Gaussian. Therefore, if $\xi_{i,j}$ is normally distributed, $P_{i,j}(Y_i\mid Y_j)$ (which represents the fluctuations in the usual scale) should follow a log-normal distribution, confirming our initial hypothesis and also connecting the surrounding fluctuations to multiplicative processes~\cite{Alves2}. Analogous to the allometric exponents, not all these distributions are independent because by knowing $P_{i,0}(Y_i\mid Y_0)$ and $P_{j,0}(Y_j\mid Y_0)$ we can also calculate $P_{i,j}(Y_i\mid Y_j)$.

\begin{figure}[!ht]
\centering
\includegraphics[scale=0.6]{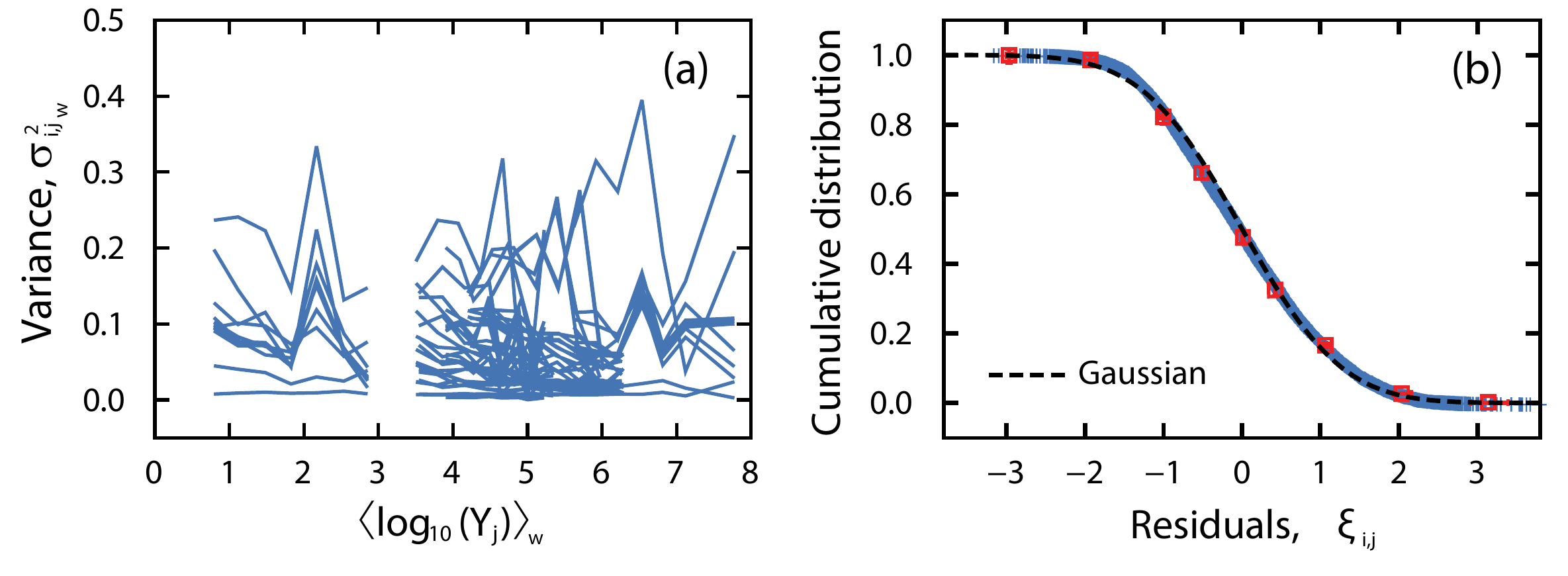}
\caption{(a) Variances $\sigma^2_{{i,j}_w}$ of the fluctuations surrounding the allometries (in base-10 logarithmic scale) for each urban indicator. It is worth noting that $\sigma^2_{{i,j}_w}$ does not display any particular form and that it can be approximated by a constant function. (b) Cumulative distributions of the normalized fluctuations $\xi_{i,j}$ surrounding the allometries (crosses). The squares are the average values over all distributions, the error bars are 95\% confidence intervals obtained via bootstrapping, and the dashed line is a standard Gaussian distribution (zero mean and unitary variance). In both plots, we have considered only the allometric relationships where the Pearson correlation $\rho_{i,j}$ is larger than $0.5$.}
\label{fig5}
\end{figure}

\section*{Summary}
We presented an extensive characterization of the connection between power-law distributions and allometries by considering 12 urban indicators from Brazilian cities. We initially verified that these 12 indicators are distributed as power laws via a rigorous statistical analysis. { Next}, we revisited the calculations of Gomez-Lievano \textit{et al.}~\cite{Lievano} for showing that, under certain hypotheses, the power-law distributions of the urban indicators can be related to the allometries between them. We empirically verified the predictions of those calculations and also whether the necessary hypotheses hold or not. In particular, we verified that the relationship between the power-law exponents and the allometric exponents is satisfied in most of the cases and that an additional condition is the quality of the allometry, measured here by Pearson correlation. We also argued that the allometric exponents are not independent of each other but, in fact, allometries between pairs of indicators ($Y_i$ and $Y_j$ with $i,j\neq0$) can be understood as a consequence of the allometric relationship of each one with the population size ($Y_0$). We further confirmed the two hypotheses underlying the calculations of Gomez-Lievano \textit{et al.}, that is, the constant behavior of the variance and the log-normal distributions of fluctuations surrounding the allometries. We thus believe that our empirical investigation contributes to consolidate the connection between power-law distributions and allometries and also reveals additional conditions underlying this relationship.

\section*{Acknowledgements}
This work has been supported by the CNPq, CAPES and Funda\c{c}\~ao Arauc\'aria (Brazilian agencies). HVR is especially grateful to Funda\c{c}\~ao Arauc\'aria for financial support under grant number 113/2013.

\end{document}